\documentstyle[prl,aps,twocolumn]{revtex}
\input prepictex
\input pictex
\input postpictex

\begin{document}
\title{
Tunneling between two Luttinger
liquids with long range interaction
}
\author{
Maura Sassetti$^{*}$ and
Bernhard Kramer\\
I. Institut f\"ur Theoretische Physik,
Universit\"at Hamburg, Jungiusstra\ss{}e 9, D--20355 Hamburg, Germany\\
(Received 31 October 1996, to be published in Physical Review B)
\vspace{3mm}
}

\author{
\parbox{14cm}{\parindent4mm\baselineskip11pt
%\begin{abstract}
{\small
The non linear charge transfer through a tunnel junction  between two
Luttinger systems is studied for repulsive, {\em finite range}
interaction between electrons on the same, $V_{11}$, and on different,
$V_{12}$, sides of the junction. Features of the Coulomb blockade
effect are observed if $V_{12}=0$. We predict a novel interaction
induced enhancement of the current if $V_{12}>0$. When $V_{12}=V_{11}$,
the current is suppressed at small bias, but the ''charging
energy'', obtained from the asymptotic behavior at high bias voltage,
vanishes.
}
%\end{abstract}
\vspace{4mm} } }

\author{ \parbox{14cm}{
 } }

\maketitle   

%\draft

Due to the repulsive interaction between the electrons, tunneling
through mesoscopic tunnel junctions is suppressed
\cite{likharev,mooijetal} for
voltages $U<U_{C}\equiv e/2C$, and temperatures $T<T_{C}\equiv
E_{C}/k_{B}$  \cite{devoret} ($k_{B}$ Boltzmann constant, $e$
elementary charge, $C$ capacitance). The quantity $E_{C}\equiv eU_{C}$
is called charging energy, the suppression of the transport 
Coulomb blockade \cite{likharev}.

In the semi-phe\-no\-menological theory \cite{ingold} the tunnel
junction is replaced by a capacitance and a tunnel resistance. An
''external'' impedance $Z(\omega )$ represents the coupling of the
tunneling particles to a reservoir of dissipative degrees of freedom.
When $Z(0)\equiv R=0$, the current-voltage characteristic $I(U)$ is
linear. For $R\neq 0$, the current is depleted for $U\ll U_{C}$. When
$U_{C}\ll U\to \infty$ $I\propto U-U_{C}$. The shift $U_{C}$ is an
important feature of the Coulomb blockade phenomenon for $R\to \infty$
which is also found in experiments \cite{devoret}.

In the quantum mechanical theory, the Hamiltonian of the system,
even using the Luttinger approximation \cite{luttinger}, becomes non
linear. For zero range repulsive interaction between the electrons, it
was shown by using renormalization group arguments
\cite{fisher,nagaosa} and by conformal field theory \cite{fendley} that
an infinitesimally small scattering barrier in a one dimensional (1D)
Luttinger liquid becomes completely insulating at zero temperature. The
current is suppressed at small voltage, $I(U)\propto U^{2/g-1}$
(interaction parameter $g<1$ for repulsive interaction). This result
was also found to the order $\Delta ^{2}$ for the model of two
Luttinger systems with $N$ ($\gg1$) transport channels connected by a tunnel
junction described by a transmission probability $\Delta$
\cite{matveev}. An effective interaction parameter $g(N)$ was
identified which tends to unity for $N\to\infty$.

In the present paper, we study a tunnel junction connecting two 1D
Luttinger systems with {\em finite} range interaction. The
correlations between the particles located at {\em different} sides of
the junction are taken into account. The current is calculated in terms
of the density-density correlation function. We recover the above
mentioned  depletion of the current at small voltages, {\em
independent} of the relative strength of the interaction between
electrons on the same, $V_{11}$ and on different sides, $V_{12}$, of
the junction. However, the latter turns out to determine the behavior
at higher bias voltages: for $V_{12}=0$ the ~~ ``Coulomb blockade'' is
found. For $V_{12}>0$, we find that the depletion of the current at
high voltage is {\em reduced}. When $V_{12}=V_{11}$, the current
approches asymptotically the non interacting, linear limit. As a
consequence, a charging energy can only be obtained for an interaction
potential with a finite, non zero range {\and} when $V_{12}<V_{11}$.
For $V_{12}>V_{11}$, though the interaction is assumed to be repulsive,
the current is {\em larger} than without interaction. 

We consider the Hamiltonian $H=H_{0} + H_{t} + H_{U}$. The unperturbed
part, $H_{0}$, describes two separate spinless Luttinger systems, 1 and
2. They are assumed to extend from $-L$ to $0$ and from $0$ to $L$
($L\to \infty$), respectively. The tunnel junction (at $x=0$) is
represented by $H_{t}$, and $H_{U}$ is the energy contributed by the
external voltage. The interaction energy is
\begin{eqnarray}
H_{\mbox{{\tiny int}}}&=&\frac{1}{2}\int_{-L}^{L}dxdy\,
\rho (x)\rho (y)\nonumber\\
&&\nonumber\\
&&\times V(|x-y|)\left[\lambda _{11}\Theta(xy) +
\lambda _{12}\Theta (-xy)\right].
\label{interaction}
\end{eqnarray}
The interaction potential $V(|x-y|)$, is assumed to introduce a
length scale, say $\alpha ^{-1}$. For a screened
potential, $V(x)\propto \exp{(- |x|/\ell)}$, $\alpha ^{-1}=\ell$. For a
Coulomb potential in 1D, $V(x)\propto (x^{2}+ d^{2})^{-1/2}$, $\alpha
^{-1}=d$ \cite{csk}. The (real and positive)
parameters $\lambda_{11}$ and $\lambda _{12}$ are introduced in order
to vary the strengths of $V_{11}\equiv \lambda _{11}V$ and
$V_{12}\equiv \lambda _{12}V$ separately. The density operator is $\rho
(x)\equiv \rho ^{(1)}(x)\Theta (-x) + \rho ^{(2)}(x)\Theta (x)$ ($\Theta
(x)$ Heavyside function).

A crucial point is that the boundary conditions \cite{fabrizio} are
such that the original Fermion fields vanish at $x=0,\pm L$. This
implies that the corresponding left and right moving parts are not
independent but $\Psi^{(j)}_{R}(x)=-\Psi ^{(j)}_{L}(-x)$, $\Psi
^{(j)}_{R}(x+2L)=\Psi ^{(j)}_{R}(x)$, and either one of the two alone
suffices to describe the system. Then, neglecting $2k_{F}$-scattering,
$\rho ^{(j)}(x)=\rho _{R}^{(j)}(x) + \rho _{R}^{(j)}(-x)$, $j=1,2$,
such that the  Fourier transform of the interaction Hamiltonian
contains terms that are non diagonal in the wave numbers. We will
see below that these lead to considerable complications in the calculation
of the non linear current-voltage relation and the charging energy.

The tunneling Hamiltonian is given by
$H_{t}\equiv H_{t}^{+} + H_{t}^{-}\equiv L \Delta (\Psi
^{(2)\dagger}_{R}(0)\Psi ^{(1)}_{R}(0)
+ \mbox{h.c.})
$.
The energy associated with the external voltage is
$H_{U}\equiv
-e\int_{-L}^{L}dx\,U(x)\rho (x)$.
We assume that the bias voltage $U(x)$ drops accross the tunnel contact
only \cite{sasskr}.

The average of the current operator
$I\equiv ie[H^{-}_{t} - H^{+}_{t}]$
is
calculated from the backward and
forward scattering rates $\gamma ^{\pm}=\int_{-\infty}^{\infty}dt\langle
H_{t}^{\pm}(t)H_{t}^{\mp}(0)\rangle$.
The average can be performed by using the phase fields $\Phi
^{(j)}(x,t)$ that are related
to the Fermion fields as usual \cite{fabrizio},
\begin{equation}
I(U)=e\left[\gamma ^{+}-\gamma ^{-}\right]=\frac{ie\Delta ^{2}}{2}
\int_{-\infty}^{\infty}dt\,\sin{(eUt)}\,e^{-W(t)}
\label{result}
\end{equation}
where
\begin{eqnarray}
W(t)&\equiv&-\langle \delta \Phi (t)\delta \Phi (0) +
(\delta \Phi (0))^{2}\rangle
\equiv\int_{0}^{\infty} d\omega \frac{J(\omega )}{\omega ^{2}}
		\times\nonumber\\
&&\nonumber\\
&&     \times \left[\left(1-\cos{(\omega t)}\right)
	\mbox{coth}\left(\frac{\beta \omega }{2}\right)
		+i\sin{(\omega t)}\right],
\label{correlation}
\end{eqnarray}
with $\delta \Phi (t)=\Phi ^{(2)}(0,t) -\Phi ^{(1)}(0,t)$ and the
spectral correlation function $J(\omega )$. This shows that the
non linear transport characteristic of the tunnel barrier is 
related to local current {\em fluctuations} represented by the phase
fields $\dot{\Phi }^{(j)}$, i.e.~the {\em dynamic} and not the sta\-tic
pro\-per\-ties of the system. This will be seen below in more detail, when
we calculate the charging energy in terms of the
interaction.

Generally, the spectral function can be decomposed
\begin{equation}
J(\omega )\equiv J_{11}(\omega )+J_{22}(\omega )+J_{12}(\omega )
+J_{21}(\omega ).
\label{generalform}
\end{equation}
It is determined by  the excitation spectrum of
the Bosonic bulk modes, $\omega (q)=v_{F}|q|/g(q)$ with $g^{-2}(q)= 1+\lambda
_{11}\hat{V}(q)/\pi v_{F}$. For small wave numbers $q$, $\omega
(q)=v_{F}|q|/g$ such that $J(\omega \to 0)$ is given by
the enhancement of the Fermi velocity $v_{F}$ with the 
interaction parameter, $g\equiv g(0)$,
that contains only $\lambda _{11}$, $
J(\omega)
=2\omega/g +
{\cal O}(\omega^{3}/ \alpha ^{2})$.
The first two terms in eq.~(\ref{generalform})
represent the contributions of the separate left and right Luttinger
systems, and are given by the expectation values $\left<\Phi
^{(j)}(0,t)\Phi ^{(j)}(0,0)\right>$. That they dominate the current at
small voltages independent of $\alpha $
and $\lambda _{12}$, is reflected in the 
behavior of $J$ for small $\omega $.

The off-diagonal terms in eq.~(\ref{generalform}) are due to the average
$\left<\Phi ^{(1)}(0,t)\Phi ^{(2)}(0,0)\right>$. They are typical vertex
contributions to the average $\langle H_{t}(t)H_{t}(0)\rangle$,
and describe correlations
between the charge fluctuations on both sides of the junction.
They influence in a characteristic way the spectral function at high
frequencies,
\begin{equation}
J(\omega \to \infty)
=2\omega \left[1 - \frac{1}{2\pi }\left(\lambda _{11}
(\omega V_{\omega })'+
2\lambda _{12}V_{\omega }\right)\right],
\label{largeomega}
\end{equation}
where $V_{\omega }\equiv \hat{V}(q=\omega /v_{F})$.

\begin{figure}
%\vspace{4cm}
%%%%%%%%%%%%%%%%%%%%%%%%%%%%%%%%%%%%%%%%%%%%%%%%%%%%%%%%%%%%%%%%%%%%%
\setcoordinatesystem units <1mm,1mm> point at 0 0 
\unitlength=1mm
\special{em:linewidth 0.4pt}
\linethickness{0.4pt}
\begin{picture}(80.00,45.00)
\plot 5 5  75 5  /
\plot 5 5  5 45  /
\plot 32 5  32 6  /
\put(3.00,44.00){\makebox(0,0)[tr]{$\frac{eR_{t}I}{\alpha v_{F}}$}}
\put(72.00,1.50){\makebox(0,0)[cb]{$eU/\alpha v_{F}$}}
\put(32.00,1.50){\makebox(0,0)[cb]{$E_{c}/\alpha v_{F}$}}
%\emline{40.00}{22.00}{5}{75.00}{42.00}{6}
%\emline{33.00}{35.00}{7}{52.00}{45.00}{8}
\setsolid
%\plot 25.00 10.00  75.00 25.00  /   
%\plot 33.00 5.00  75.00 29.00  / 
\plot 47.00 14.00  75.00 30.00  /
\setquadratic
\plot 5.00 5.00  26.50 6.75  47.00 14.00  /
\setdashes <1mm>
\plot 5.00 5.00  19.00 6.00  25.00 11.00  30.00 16.50  40.00 23.00  /
\setlinear
\plot 40.00 23.00  75.00 43.00  /
\setquadratic
\setdashes <0.5mm>
\plot 5.00 5.00  10.50 6.00  14.00 12.00  17.00 22.00  25.00 30.00  /
\setlinear
\plot 25.00 30.00  51.00 45.00  /
\setdashes <2mm>
\setlinear
\plot 5.00 5.00  75.00 45.00  /
\setdots
\plot 32.00 5.00  47.00 14.00  /
\end{picture}
%%%%%%%%%%%%%%%%%%%%%%%%%%%%%%%%%%%%%%%%%%%%%%%%%%%%%%%%%%%%%%%%%%%%%%
\caption[I(U)]{The zero temperature current-voltage characteristic,
schematically,
of a tunnel junction (tunneling resistance $R_{t}$)
connecting two semi infinite Luttinger liquids for
various ratios of the interaction strengths between electrons on 
different, $\lambda _{12}$, and the same, $\lambda _{11}$,
sides of the junction. Full line: $\lambda _{12}/\lambda _{11}=0$; short
dashes: $\lambda _{12}/\lambda _{11}<1$; 
dots: $\lambda _{12}/\lambda _{11}>1$; long dashes: no interaction.
}
\label{I(U)}
\end{figure}
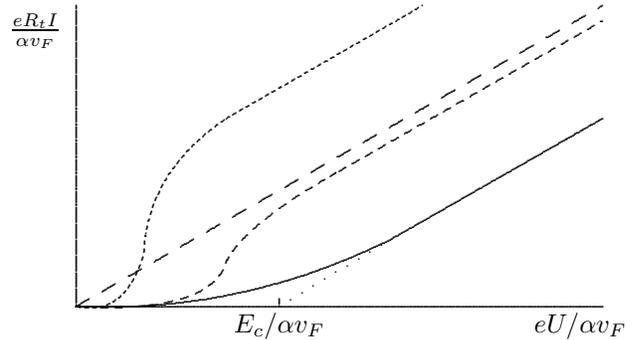

Since the interaction range is assumed to be finite in position space,
the second term on the right hand side of eq.~(\ref{largeomega})
vanishes for $\omega \to \infty$ such that only the contribution of the
free, non interacting electrons survives, namely $J(\omega )=2\omega$.
The presence of the term $\propto \lambda _{12}$ on the right hand side
reflects the above mentioned fluctuation correlations. We will see
below in detail that the latter counteract the suppression of the
current induced by the coupling to the bulk modes and eventually lead
to a vanishing of the charging energy if $\lambda _{11}=\lambda
_{12}$. If $\lambda _{12}>\lambda _{11}$ the current at high bias
voltage becomes even enhanced {\em above} the value  without
interaction. The behavior of the current-voltage characteristic at zero
temperature is shown schematically in Fig.~\ref{I(U)}.

The connection with the earlier, semi classical results 
\cite{ingold}
is established by rewriting
$I(U)$ as
\begin{eqnarray}
I(U)&=&\frac{1-e^{-\beta eU}}{eR_{t}}
	\int_{-\infty}^{\infty}dE\int_{-\infty}^{\infty}dE'
	f(E)[1-f(E')]\nonumber\\
	&&\nonumber\\
	&&\qquad\qquad\qquad\qquad\qquad\times P(E+eU-E'),
\label{iofu}
\end{eqnarray}
with the Fermi function $f(E)$ and 
\begin{equation}
P(E)=\frac{1}{2\pi }\int_{-\infty}^{\infty}dt e^{iEt}e^{-[W(t)-W_{g=1}(t)]}.
\label{p(e)}
\end{equation}
Only in the semi classical approach, the latter function plays the role
of a probability density for a bulk excitation of energy $E$. In the
present microscopic model, it can become negative. This is a result of
the extraction of the Fermi factor $f(1-f)$ in the integrand in
eq.~(\ref{iofu}) which is artificial, though formally correct, for
the present
interacting system. The tunnel resistance
$R_{t}\equiv 2\omega _{c}^{2}/e^{2}\Delta ^{2}\pi$ contains the cutoff
frequency $\omega _{c}$ which serves to regularize the Luttinger model.

The result for zero temperature,
\begin{equation}
I(U)=\frac{1}{eR_{t}}\int_{0}^{eU}dE(eU -E)P(E), \label{zerot}
\end{equation}
can be used to identify the parameters in the semi
classical approach in terms of the microscopic model.
The function that corresponds to the impedance in the semi classical
theory is $Z(\omega )\equiv J(\omega )/\omega -2$. For small bias,
$eU\ll \alpha v_{F}$, only the behavior of $P(E)$ for small energies is
important. It is determined by the diagonal contributions in
eq.~(\ref{generalform}) and always positive. The electrons loose
energy via dissipation into the bulk modes. We find $I(U)\approx
(U/R_{t})(eU/\alpha v_{F})^{2/g-2}$, which yields the dissipative resistance
\cite{matveev} $R/R_{K}\equiv Z(0)= 2(g^{-1}-1)$ ($R_{K}=h/e^{2}$ von
Klitzing constant). This shows that the elementary excitations of the
Luttinger system play here the role of the dissipative Bosonic degrees
of freedom of the ''electromagnetic environment''.

At high bias voltage, $eU>\alpha v_{F}$, $J_{12}$ and $J_{21}$ become
dominant. They counteract the two dissipative diagonal terms $J_{11}$
and $J_{22}$. While the latter always supress, the former tend to {\em
enhance}  transport. Formally, this can be seen from $P(E\to \infty$).
We start by calculating the second derivative of the current-voltage
relation, eq.~(\ref{zerot}) $I''(U)=eP(eU)/R_{t}$. For very large $U$,
$P(eU)=Z(eU)/eU$. If we make the (not very restrictive) assumption that
 for large wave numbers $\hat{V}(q)\propto q^{-2}$, we have  $\pi
Z(\omega \to \infty)=2V_{\omega }(\lambda _{11}/2-\lambda _{12})$ which
changes sign when  $\lambda _{12}>\lambda _{11}/2$. When the coupling
between electrons left and right of the junction exceeds a certain
strength, the curvature of the current-voltage curve changes from
positive to negative. This occurs for voltages that correspond to the
intrinsic length scale $\alpha ^{-1}$.

For $U\gg \alpha v_{F}$, only short times contribute to the integrand
of eq.~(\ref{p(e)}), the current, eq.~(\ref{zerot}), becomes
$
I(U)=
R_{t}^{-1}(U-E_{C}/e) +
{\cal O}(U^{-1})
$ with the charging energy
\begin{equation}
E_{C}\equiv\int_{0}^{\infty}d\omega Z(\omega )=V(x=0)(\lambda _{11}-\lambda _{12}).
\label{chargingenergy}
\end{equation}
The derivation of this result
is far from trivial.
The starting point is the relation between $W(t)$, eq.~(\ref{correlation}),
and the fluctuation correlation function $G_{kk'}(t)$
\begin{eqnarray}
F(t)&\equiv &-i\Theta (t)
\left<\left[\delta \Phi (0,t),\delta \Phi (0,0)\right]\right>\nonumber\\
&&\nonumber\\
&\equiv&\frac{4\pi }{L}\sum_{k,k'>0}\frac{1}{kk'}G_{kk'}(t).
\end{eqnarray}
In $W(t)$, the average
$\left<\delta \Phi (0,t)\delta \Phi (0,0)\right>$
enters. It can be expressed via the dissipation-fluctuation theorem in
terms of the imaginary part of the Fourier transform of $F(t)$. This
leads eventually to the identity
\begin{equation}
J(\omega )=-\frac{1}{\pi }\omega ^{2}{\cal I}m\hat{F}(\omega ).
\end{equation}

The set of equations of motion for the correlation function can be
closed for our model. The resulting linear integral equation is  used
to expand $\hat{G}_{kk'}(\omega )$ in powers of
$\hat{G}^{(0)}_{k}(\omega )B_{kk'}(\omega )$, where $\hat{G}^{(0)}_{k}\equiv
\omega ^{2}/(\omega ^{2}-\omega ^{2}_{k})$ is the unperturbed
correlation function. It contains the excitation spectrum of the bulk
modes. The function $B_{kk'}$
contains the off-diagonal part of the interaction,
$B_{kk'}(\omega )\equiv -
(\lambda _{11}+\lambda _{12}){\cal V}(k,k')kk'/\omega ^{2}\pi L$ with
\begin{equation}
{\cal V}(k,k')\equiv -\frac{{\cal P}}{\pi }
\int_{-\infty}^{\infty}\,
dq\frac{\hat{V}(q)q^{2}}{(q^{2}-k^{2})(q^{2}-k'^{2})}.
\end{equation}

By performing first the $\omega
$-integration (cf. eq.~(\ref{chargingenergy}))
and then the summations over the wave numbers, one can
show that in the integral over the imaginary part of $\hat{F}$
only the term linear in
$B_{kk'}$ contributes. All higher order contributions vanish,
due to exact sum rules. The calculation leads eventually to 
\begin{eqnarray}
E_{C}&\equiv &-2\int_{0}^{\infty}\,d\omega \left[\frac{2}{L}\sum_{k,k'}
\frac{\omega {\cal I}m\hat{G}_{kk'}(\omega )}{kk'}+1\right]\nonumber\\
&&\nonumber\\
&=&E_{C}^{0}-\frac{2(\lambda _{11}+\lambda _{12})}{\pi ^{2}}
\int_{0}^{\infty}dk\,dk'\,{\cal V}(k,k').
\end{eqnarray}
The first term in this expression is the result
obtained without taking
into account the off-diagonal interaction
terms, eq.~(\ref{interaction}) \cite{sassetal},
$E_{C}^{0}=2\lambda _{11}V(x=0)$. The second
can be further evaluated and gives
$-(\lambda _{11}+\lambda _{12})V(x=0)$ such
that finally the above result
eq.~(\ref{chargingenergy}) is obtained.

There are several comments which have to be made at this point. When
the interaction between electrons left and right of the tunnel junction
is omitted, $\lambda _{11}=1$, $\lambda _{12}=0$ the preliminary result
obtained previously  \cite{sassetal} is recovered apart from a factor
1/2 which is due to discarding the influence of the boundary conditions
on the fields. The latter influence is negligible for small frequencies
\cite{fabrizio}. For high frequencies, the boundary conditions
influence quantitatively the result for the charging energy, though the
qualitative behavior with the strength and the range of the interaction
is the same as before. When the interaction of electrons in the left
and the right part of the system is switched on, $\lambda _{12}\neq 0$,
the charging energy is {\em reduced} as compared to the previous case.
For a given voltage, the current is increased, due to the presence of
the additional, (formally) repulsive electron-electron term. In
principle, when the interaction between electrons in the left and the
right part of the system dominates, $\lambda _{11}<\lambda _{12}$, the
current increases even to values above that of the non-interacting
system, $I(U)>U/R_{t}$.

At the first glance, this seems counter-intuitive. How can a
repulsive interaction term {\em increase} the current? The puzzle can
be solved by noting that this happens at {\em high voltages}. For
sufficiently small bias, in the regime dominated by the coupling to
the bulk modes, the current remains depleted well below $U/R_{t}$. When
the bias voltage is sufficiently high, net charges with opposit signs
are introduced left and right of the tunnel junction. This is
also suggested by the model of the ''Landauer dipole'' at an impurity
in the presence of a stationary current \cite{zwerger}. As a
consequence, the net interaction between the left and right
Luttinger liquids becomes attractive and tends to decrease the charge
difference -- the bias voltage -- via additional (tunnel) current. This
cannot happen in the semi classical model since there the
quantum processes are introduced into the model ''by hand'', and
not inherently incorporated into the theory. 

Insisting nevertheless on assigning a {\em capacitance} to the tunnel
junction leads to a dependence of the latter on the bias voltage
\cite{delsing}: for  small $U$, the capacitance would be finite, due to
the suppression of the current well below the value of the system
without interaction. When increasing the bias voltage the correlations
between the electrons on the left and the right of the junction
increase the current, i.e.~ decrease the effective charging energy
such that the ''capacitance'' would increase to infinity when $\lambda
_{12}\to \lambda _{11}$. This shows its {\em dynamical} origin \cite{csk}.

A similar result is obtained when considering a potential
barrier in a Luttinger liquid, where the interaction between
electrons left and right of the barrier is taken into account
automatically, $\lambda _{11}=\lambda _{12}$.

Experimentally, it is very difficult to
observe Coulomb blockade for a single tunnel junction between two
metallic wires \cite{delsing,1junctionexp} due to the presence of a
large shunt capacitance. By adding more junctions in series, this shunt
capacitance is reduced due to the small capacitances of the other
junctions.  Our above result offers a {\em
microscopic} interpretation: adding junctions in series reduces
the interaction between electrons in the left and right leads. As a
consequence, the observed charging energy is increased.

A direct experimental test could be performed by using quantum wires
based on semiconductor hetero structures and adjusting the electron
density such that only one subband is occupied. In the presence of a
gate across the region of the tunnel barrier, the interaction between
the leads on the left and the right hand side of the barrier could be
locally screened such that the charging energy should appear.

In summary, we obtained the non linear current-voltage characteristic
for a model of two 1D quantum wires of interacting electrons connected
by a tunnel junction. The features of the Coulomb blockade phenomenon
were identified. The
charging energy was determined as a function of the parameters of the
interaction potential. We showed that it is  crucial in this
microscopic model that the interaction, besides being of {\em finite range},
must {\em not} be too strong between electrons on different sides of the
junction in order to produce a charging energy.

Discussions with Det\-lef Heit\-mann, Wolf\-gang Han\-sen, Klaus von
Klitzing and Leo Kouwenhoven con\-cer\-ning experi\-ments are
gra\-te\-ful\-ly ack\-now\-led\-ged. This work has been supported by
the EU via HCM and TMR prog\-rammes, contracts CHRX-CT93 0136,
CHRX-CT94 0464, FMRX-CT96 0042, and by the
Deut\-sche For\-schungs\-ge\-mein\-schaft via the
Gra\-du\-ier\-ten\-kol\-leg ''Physik nano\-struk\-tu\-rier\-ter
Fest\-k\"or\-per''.

\vspace{3mm}
$^{*}$ {\small on leave of absence from Istituto di Fisica di Ingegneria, INFM,
Universit\`a di Genova, Italy}

\end{document}